\documentclass[preprint,12pt]{elsarticle}

\usepackage{lineno,hyperref} 

\usepackage{longtable}
\usepackage{color}
\usepackage{lscape}
\usepackage{graphicx}

\begin{document}

\begin{frontmatter}

\title{Spectroscopic and Electric Properties of the TaO$^+$ Molecule Ion for the Search of New Physics: A Platform for Identification and State Control}

\author[address1]{Ayaki Sunaga\corref{corresponding}}
\ead{sunagaayaki@gmail.com}
\author[address2]{Timo Fleig}
\ead{timo.fleig@irsamc.ups-tlse.fr}

\address[address1]{Institute for Integrated Radiation and Nuclear Science, Kyoto University, Osaka
590-0494, Japan}
\address[address2]{Laboratoire de Chimie et Physique Quantiques, FeRMI, Universit{\'e} Paul Sabatier Toulouse III, 118 route de Narbonne, F-31062 Toulouse, France}


\cortext[corresponding]{Corresponding author}


\begin{abstract}
The TaO$^+$ cation is an attractive molecular species to search for parity- and time-reversal-violating interactions, in particular of hadronic origin. For the spectroscopic detection and preparation of TaO$^+$ cation in a desired state detailed knowledge of spectroscopic and electric properties in excited states is essential information. In this work we present spectroscopy constants for TaO$^+$ in the electronic ground and 29 excited states calculated with  relativistic configuration interaction theory. The equilibrium bond lengths ($R_\mathrm{e}$), harmonic vibrational frequencies ($\omega_\mathrm{e}$), transition dipole moments (TDM), vertical excitation energies and static molecular dipole moments (PDM) are summarized. We include a detailed characterization of all electronic states in terms of their spinor occupations. This work supports the realization of experiments using TaO$^+$ ions to search for new physics beyond the standard model of elementary particles.
\end{abstract}


\end{frontmatter}


\section{Introduction}

In the search for atomic-scale manifestations of charge-parity violation (CPV)
\cite{Kobayashi} beyond that known and already incorporated into the Standard Model (SM) of
elementary particles, high-precision measurements on molecular ions 
\cite{HfF+_EDM_PRL2017,Cornell_TrapEDM_2020} have become a competitive alternative to measurements with electrically neutral systems
that so far yield the strongest constraints on fundamental leptonic and
semi-leptonic CPV parameters
\cite{ACME_ThO_eEDM_nature2018,hudson_hinds_YbF2011}. In addition, the
upper bounds on the measured ${\cal{P-}}$ and ${\cal{T-}}$violating (where ${\cal{T}}$ stands for time-reversal) molecular electric dipole moments (EDM) also provide constraints on CPV in the hadron sector through the nuclear magnetic quadrupole moment (MQM)
\cite{Sushkov_Flambaum_Khriplovich1984}
which itself originates in the quantum chromodynamics CPV parameter $\theta$ \cite{Veneziano_Witten_QCD-PTodd1980}, the EDMs of the $u$ and $d$
quarks, and via chromo-EDMs \cite{Gunion_Wyler_CEDM_NEDM_1990}. 

In earlier work \cite{PhysRevA.95.022504} the tantalum oxide cation (TaO$^+$)
has been characterized as a very promising molecular ionic probe for hadronic
CPV {\it{via}} the nuclear MQM. The major findings in that work have been
the sizable nuclear-MQM-electronic-field interaction constant of 
$W_M = 0.45$ [$\frac{10^{33} {\rm{Hz}}}{e\, {\rm{cm}}^2}$] and the 
fact that the electronic ground state is characterized by a ${^3\Delta_1}$ 
term. Such a state is used in ongoing EDM experiments with the hafnium
flouride cation (HfF$^+$) \cite{HfF+_EDM_PRL2017} and near-future EDM
experiments with thorium flouride cation (ThF$^+$) \cite{Zhou2019JMS_ThF+,cornell_ThF+_BAPS_2021,Ng2022PRA_ThF+},
where in the latter ${^3\Delta_1}$ is also the electronic ground state
\cite{ThF+_NJP_2015,Gresh2016JMS}. Therefore, experiments on such systems may attain coherence times on the scale of many seconds \cite{Cairncross_Ye_NatPhys2019}.
Moreover, suitable nuclear isotopes with strong deformation may give rise to large nuclear MQMs. Comparative estimates have shown that {$^{181}$Ta} potentially has among the largest nuclear MQMs of heavy target atoms used
in EDM experiments \cite{Flambaum_DeMille_Kozlov2014,Lackenby2018PRA_MQM}. The effect of the mixing of $^3\Delta_1$ and $^3\Delta_{-1}$ states by the tensor parity non-conserving (PNC) interaction has also been reported \cite{Penyazkov2022CPL_PNC}.

In practical experiments molecular species are often created and/or identified through transitions with electronically excited states \cite{CarrDeMille_NJP2009,Cornell_MolIons_Science2013,Ni_HfF+_JMS2014}. The identification and characterization
as well as making reliable predictions
for energies and (transition) properties of such excited molecular states is, therefore, of value in conceptualizing and carrying out a high-precision molecular experiment. 

The string-based generalized-active-space configuration interaction (GASCI) method \cite{knecht_luciparII,fleig_gasci2} is a versatile and reliable tool for addressing the complicated electronic structure of diatomic molecules in excited states. This is a particular asset in transition-metal molecules like the present TaO$^+$ cation that are especially challenging due to their open $d$ shell and the ensuing strong multi-configurational character of many excited states. 
We use this method in the present work to address a large number of excited states of TaO$^+$ covering an energy window of about $5$ eV above the electronic ground state. 

The paper is structured as follows: Section \ref{sec:theory} documents the computational methodology employed in this study. The details of the reference spinors and GASCI are described. Next, we characterize the electronic ground and excited states on the basis of the spectroscopic constants, transition dipole moments (TDM), molecular permanent dipole moments (PDM), and the electronic configuration in section \ref{sec:results}. We map out the strongest transitions among the complete ensemble of excited states
in the respective energy window which may serve to design a pathway to reach the science state, $^3\Delta_1$. The results of the low-energy excited states in the previous work \cite{PhysRevA.95.022504} confirm the validity of this calculation. Finally, we summarize and conclude (section \ref{sec:conclusion}).

\section{Computational details}\label{sec:theory}
\subsection{Generalities}
All calculations are carried out with the DIRAC15 software package \cite{DIRAC15,Saue2020JCP} with revisions 039d222. The (spin-orbit) Dirac-Coulomb Hamiltonian with the (SS\textbar SS) type integral approximated as the classical repulsion\cite{Visscher1997TCA} are employed. Dyall.4zp basis sets without polarization functions \cite{Dyall2004TCA_5d,Dyall2009TCA_5d,Dyall2016TCA_1-3} are used in uncontracted form (Ta:34s30p19d12f; O:18s10p). The small components of the basis sets are generated based on restricted kinetic balance \cite{Stanton1984JCP}. The Gaussian-type finite-nuclear model \cite{Visscher1997ADNDT} was employed for the nucleus-electron interaction.

We used the KR-CI module \cite{Fleig2003JCP,Knecht2010JCP} (which incorporates the GASCI method) to calculate the spectroscopic constants and molecular properties \cite{Knecht_thesis,Denis2015NJP} based on relativistic CI theory. If not stated otherwise, we employed MR$^{+T}_{10}$-CISD(8) with $n=0$ and 2.5 $E_H$ truncation,
where the CISD calculation is defined with eight electrons in the active space and three holes in O's $2p$-dominant spinors. 
Virtual truncations of 5.0 $E_H$ and 10.0 $E_H$ were tested for comparison.
Details of the terminology of the KR-CI method and the active space model are
given in Ref. \cite{Fleig2016PRA_TaN}. The reference spinors have been obtained through the Dirac-Coulomb average-of-configuration Hartree-Fock (AOC-HF) method \cite{Thyssen_thesis}. The details of AOC-HF method are discussed in the next section \ref{sec:MP}.

\subsection{Spinor basis set}\label{sec:MP}
In the case of molecules including $d$ and $f$ block elements, the choice of the reference DCHF spinors significantly affects the electronic excited states  \cite{fleig_nayak_eEDM2013,Denis-Fleig_ThO_JCP2016}. The Mulliken populations of the following two references are summarized in Table \ref{tbl:MP}: i) two electrons are evenly distributed in $5d$ and $6s$ spinors, which is the same in the previous works of TaO$^+$ \cite{PhysRevA.95.022504} and TaN \cite{Fleig2016PRA_TaN}, referred to as DCHF\_2in12. ii) two electrons are evenly distributed in Ta's $6s$, $5d$, $6p$, and $7s$ spinors, referred to as DCHF\_2in20. In the case of DCHF\_2in12, the no. 46-49 molecular spinors predominantly consist of Ta's $6p$ and $7s$ spinors. Meanwhile, the $6p$ and $7s$ spinors are mixed with $5d$ in the no. 46-49 molecular spinors of DCHF\_2in20 model.
The inclusion of $6p$-occupied configuration is also essential for the accurate calculation of the TDM, where the excitation $6s\rightarrow6p$ occurs, as shown in the section \ref{sec:spec}. Although the DCHF\_2in12 would be good for the ground state and low-energy excited states that are characterized by $6s$ and $5d$ spinor occupations, we use the DCHF\_2in20 model to calculate high-energy excited states with similar accuracy.

\section{Results and discussion}\label{sec:results}
The dataset of this work is available via the zenodo repository \cite{zenodo:dataset}.

\begin{table}[ht]                                                                   
\begin{center}                                                                  
\caption{Characterization of the Kramers pairs in active space and Ta's $7p$-dominant spinors in terms of orbital angular momentum projection. Occupation number (occ.),  Mulliken population (MP), projection quantum number (${m_j}$) of total angular momentum $j$, and spinor energy ($\varepsilon$) at 3.1609 bohr internuclear distance. Two electrons in $6s5d$ is referred to as DCHF\_2in12 and two electrons in $6s5d7s6p$ is referred to as DCHF\_2in20. The no. 37-39 spinors are O's $2p$-dominant ones \cite{PhysRevA.95.022504}}\label{tbl:MP}
\begin{tabular}{c|cccc|cccc}
\hline \hline
\multicolumn{1}{c}{} & \multicolumn{4}{c}{DCHF\_2in12} & \multicolumn{4}{c}{DCHF\_2in20}    \\
No. & occ. & $|{m_{j}}|$ & $\varepsilon(E_H)$  & MP atom($l_{\lambda}$) & occ. & $|{m_{j}}|$ & $\varepsilon(E_H)$& MP atom($l_{\lambda}$) \\
\hline & & & & $78 \%\;  \mathrm{Ta}(s)$ & & & & $79 \%\; \mathrm{Ta}(s)$ \\
40 & $2 / 12$ & $1 / 2$ & $-0.590$ & $24 \%\; \mathrm{Ta}\left(d_{\delta}\right)$ & $2 / 20$ & $1 / 2$ & $-0.614$ & $25 \%\; \mathrm{Ta}\left(d_{\delta}\right)$ \\
& & & & $-8 \%\; \mathrm{Ta}\left(d_{\sigma}\right)$ & & & & $-8 \%\; \mathrm{Ta}\left(d_{\sigma}\right)$ \\
\hline 41 & $2 / 12$ & $3 / 2$ & $-0.575$ & $100 \%\; \mathrm{Ta}\left(d_{\delta}\right)$ & $2 / 20$ & $3 / 2$ & $-0.599$ & $99 \%\; \mathrm{Ta}\left(d_{\delta}\right)$ \\
\hline 42 & $2 / 12$ & $5 / 2$ & $-0.575$ & $100 \%\; \mathrm{Ta}\left(d_{\delta}\right)$ & $2 / 20$ & $5 / 2$ & $-0.597$ & $100 \%\; \mathrm{Ta}\left(d_{\delta}\right)$ \\
\hline & & & & $20 \%\; \mathrm{Ta}\left(p_{\pi}\right)$ & & & & $21 \%\; \mathrm{Ta}\left(p_{\pi}\right)$ \\
43 & $2 / 12$ & $1 / 2$ & $-0.506$ & $62 \%\; \mathrm{Ta}\left(d_{\pi}\right)$ & $2 / 20$ & $1 / 2$ & $-0.532$ & $50 \%\; \mathrm{Ta}\left(d_{\pi}\right)$ \\
& & & & $12 \%\; \mathrm{O}\left(p_{\pi}\right)$ & & & & $14 \%\; \mathrm{O}\left(p_{\pi}\right)$ \\
\hline & & & & $16 \%\; \mathrm{Ta}\left(p_{\pi}\right)$ & & & & $18 \%\; \mathrm{Ta}\left(p_{\pi}\right)$ \\
44 & $2 / 12$ & $3 / 2$ & $-0.492$ & $68 \%\; \mathrm{Ta}\left(d_{\pi}\right)$ & $2 / 20$ & $3 / 2$ & $-0.519$ & $68 \%\; \mathrm{Ta}\left(d_{\pi}\right)$ \\
& & & & $14 \%\; \mathrm{O}\left(p_{\pi}\right)$ & & & & $16 \%\; \mathrm{O}\left(p_{\pi}\right)$ \\
\hline & & & & $6 \%\; \mathrm{Ta}(s)$ & & & & $7 \%\; \mathrm{Ta}(s)$ \\
45 & $2 / 12$ & $1 / 2$ & $-0.432$ &   $39 \%\; \mathrm{Ta}\left(p_{\sigma}\right)$ & $2 / 20$ & $1 / 2$ & $-0.453$ & $43 \%\; \mathrm{Ta}\left(p_{\sigma}\right)$ \\
& & & & $31 \%\; \mathrm{Ta}\left(d_{\sigma}\right)$ & & & &$28 \%\; \mathrm{Ta}\left(d_{\sigma}\right)$ \\
& & & & $11 \%\; \mathrm{O}\left(p_{\sigma}\right)$ & & & & $11 \%\; \mathrm{O}\left(p_{\sigma}\right)$ \\
\hline  46 & 0 & $1/2$ & $-0.108$ & $93 \%\; \mathrm{Ta}\left(p_{\pi}\right)$ & $2 / 20$ & $1 / 2$ & $-0.365$ & $80 \%\; \mathrm{Ta}\left(p_{\pi}\right)$  \\
& & & & & & & & $12 \%\; \mathrm{Ta}\left(d_{\pi}\right)$ \\
\hline  47 & 0 & $3 / 2$ & $-0.106$ & $96 \%\; \mathrm{Ta}\left(p_{\pi}\right)$ & $2 / 20$ & $3 / 2$ & $-0.355$ & $84 \%\; \mathrm{Ta}\left(p_{\pi}\right)$  \\
& & & & & & & & $12 \%\; \mathrm{Ta}\left(d_{\pi}\right)$ \\
\hline 48 & 0 & $1 / 2$ & $-0.078$ & $89 \%\; \mathrm{Ta}(s)$  & $2 / 20$ & $1 / 2$ & $-0.238$ &  $85 \%\; \mathrm{Ta}(s)$ \\
& & & &$7 \%\; \mathrm{Ta}\left(p_{\sigma}\right)$ & & & & $8 \%\; \mathrm{Ta}\left(p_{\sigma}\right)$ \\
& & & & & & & &$15 \%\; \mathrm{Ta}\left(d_{\delta}\right)$  \\
& & & & & & & &$-8 \%\; \mathrm{Ta}\left(d_{\sigma}\right)$  \\
\hline  49 & 0 & $1 / 2$ & $-0.064$ & $92 \%\; \mathrm{Ta}\left(p_{\sigma}\right)$ & $2 / 20$ & $1 / 2$ & $-0.223$ & $12 \%\; \mathrm{Ta}\left(s\right)$ \\
& & & & & & & &$75 \%\; \mathrm{Ta}\left(p_{\sigma}\right)$  \\
& & & & & & & &$14 \%\; \mathrm{Ta}\left(d_{\sigma}\right)$   \\
\hline  50 & 0 & $1 / 2$ & $-0.045$ & $100 \%\; \mathrm{Ta}\left(p_{\pi}\right)$ & 0 & $1 / 2$ & $-0.060$ & $100 \%\; \mathrm{Ta}\left(p_{\pi}\right)$ \\
\hline  51 & 0 & $3 / 2$ & $-0.043$ & $99 \%\; \mathrm{Ta}\left(p_{\pi}\right)$ & 0 & $3 / 2$ & $-0.058$ & $100 \%\; \mathrm{Ta}\left(p_{\pi}\right)$ \\
\hline  52 & 0 & $3 / 2$ & $0.014$ & $97 \%\; \mathrm{Ta}\left(d_{\delta}\right)$ & 0 & $1 / 2$ & $-0.024$ & $88 \%\; \mathrm{Ta}\left(p_{\sigma}\right)$ \\
\hline \hline
\end{tabular}
\end{center}                                                                    
\end{table}

\subsection{Active spinor space}
The values of $T_v$ are sensitive to the number of holes in the active space, while the size of the virtual space does not affect it so much. The size of the virtual space does not affect the transition energy $T_v$ except for high-energy excited states that are close to 5 eV above the electronic ground state (Table \ref{tbl:active}). Even the order of the $^{2S+1}\Lambda_{\Omega}$ terms is left unchanged by the virtual cutoff. On the other hand, the number of correlated electrons on oxygen affects $T_v$ values even for the low-energy excited states: for instance, the change in excitation energy for the first ${}^{1} \Sigma_{0}$ at the 2-hole and 3-hole levels is about 0.05 eV (Table \ref{tbl:core}). Due to the significant dependency on the number of the correlated electrons the results for the states that are close to 5 eV may be unreliable. Calculations at the 4-hole level may thus be warranted for highly accurate results, but they are very costly. We chose to employ the 3-hole model and 2.5 $E_H$ truncation level for the final calculations. We did not correlate oxygen’s $2s$-dominant spinor (-1.425 $E_H$) because it is energetically separated from the $2p$-dominant spinor (-0.755 $E_H$) \cite{PhysRevA.95.022504}.

\begin{table}[ht]                                               
\begin{center}                                                  
\caption{Calibration of the truncation of the active space based on MR$_{10}$-CISD(8) model (i.e. two holes in O's $2p$-dominant spinors) on the transition energy ($T_v(R);R=3.1609$ bohr).}\label{tbl:active}
\begin{tabular}{ccc|ccc}
\hline\hline
\multicolumn{3}{c}{$T_v$ (eV)} &\multicolumn{3}{c}{$^{2S+1}\Lambda_{\Omega}$} \\
$2.5$ ($E_H$) & 5.0 ($E_H$) &10.0 ($E_H$)& $2.5$ ($E_H$) & 5.0 ($E_H$) & 10.0 ($E_H$) \\
\hline 
 $0.00$ & $0.00$ & $0.00$ & ${ }^{3} \Delta_{1}$ & ${ }^{3} \Delta_{1}$ & ${ }^{3} \Delta_{1}$ \\
 $0.16$ & $0.16$ & $0.16$ & ${ }^{3} \Delta_{2}$ & ${ }^{3} \Delta_{2}$ & ${ }^{3} \Delta_{2}$ \\
 $0.36$ & $0.37$ & $0.37$ & ${ }^{1} \Sigma_{0}$ & ${ }^{1} \Sigma_{0}$ & ${ }^{1} \Sigma_{0}$ \\
 $0.39$ & $0.40$ & $0.40$ & ${ }^{3} \Delta_{3}$ & ${ }^{3} \Delta_{3}$ & ${ }^{3} \Delta_{3}$ \\
 $1.16$ & $1.15$ & $1.15$ & ${ }^{3} \Sigma_{0}$ & ${ }^{3} \Sigma_{0}$ & ${ }^{3} \Sigma_{0}$ \\
 $1.23$ & $1.22$ & $1.22$ & ${ }^{3} \Sigma_{1}$ & ${ }^{3} \Sigma_{1}$ & ${ }^{3} \Sigma_{1}$ \\
 $1.34$ & $1.34$ & $1.34$ & ${ }^{1} \Delta_{2}$ & ${ }^{1} \Delta_{2}$ & ${ }^{1} \Delta_{2}$ \\
 $1.86$ & $1.85$ & $1.85$ & ${ }^{1} \Gamma_{4}$ & ${ }^{1} \Gamma_{4}$ & ${ }^{1} \Gamma_{4}$ \\
 $2.15$ & $2.14$ & $2.14$ & ${ }^{3} \Phi_{2}$ & ${ }^{3} \Phi_{2}$ & ${ }^{3} \Phi_{2}$ \\
 $2.17$ & $2.18$ & $2.18$ & ${ }^{3} \Pi_{0}$ & ${ }^{3} \Pi_{0}$ & ${ }^{3} \Pi_{0}$ \\
 $2.20$ & $2.21$ & $2.21$ & ${ }^{3} \Pi_{0}$ & ${ }^{3} \Pi_{0}$ & ${ }^{3} \Pi_{0}$ \\
 $2.23$ & $2.23$ & $2.23$ & ${ }^{3} \Pi_{1}$ & ${ }^{3} \Pi_{1}$ & ${ }^{3} \Pi_{1}$ \\
 $2.41$ & $2.42$ & $2.42$ & ${ }^{3} \Pi_{1}$ & ${ }^{3} \Pi_{1}$ & ${ }^{3} \Pi_{1}$ \\
 $2.43$ & $2.42$ & $2.42$ & ${ }^{1} \Sigma_{0}$ & ${ }^{1} \Sigma_{0}$ & ${ }^{1} \Sigma_{0}$ \\
 $2.53$ & $2.53$ & $2.53$ & ${ }^{3} \Phi_{3}$ & ${ }^{3} \Phi_{3}$ & ${ }^{3} \Phi_{3}$ \\
 $2.57$ & $2.58$ & $2.58$ & ${ }^{3} \Pi_{2}$ & ${ }^{3} \Pi_{2}$ & ${ }^{3} \Pi_{2}$ \\
 $2.79$ & $2.78$ & $2.78$ & ${ }^{3} \Pi_{0}$ & ${ }^{3} \Pi_{0}$ & ${ }^{3} \Pi_{0}$ \\
 $2.81$ & $2.80$ & $2.80$ & ${ }^{3} \Pi_{0}$ & ${ }^{3} \Pi_{0}$ & ${ }^{3} \Pi_{0}$ \\
 $2.83$ & $2.83$ & $2.83$ & ${ }^{3} \Pi_{2}$ & ${ }^{3} \Pi_{2}$ & ${ }^{3} \Pi_{2}$ \\
 $2.91$ & $2.91$ & $2.91$ & ${ }^{1} \Pi_{1}$ & ${ }^{1} \Pi_{1}$ & ${ }^{1} \Pi_{1}$ \\
 $2.93$ & $2.92$ & $2.92$ & ${ }^{3} \Phi_{4}$ & ${ }^{3} \Phi_{4}$ & ${ }^{3} \Phi_{4}$ \\
 $3.53$ & $3.54$ & $3.54$ & ${ }^{1} \Phi_{3}$ & ${ }^{1} \Phi_{3}$ & ${ }^{1} \Phi_{3}$ \\
 $3.74$ & $3.75$ & $3.75$ & ${ }^{3} \Delta_{1}$ & ${ }^{3} \Delta_{1}$ & ${ }^{3} \Delta_{1}$ \\
 $3.90$ & $3.90$ & $3.90$ & ${ }^{3} \Delta_{2}$ & ${ }^{3} \Delta_{2}$ & ${ }^{3} \Delta_{2}$ \\
 $3.92$ & $3.93$ & $3.93$ & ${ }^{1} \Pi_{1}$ & ${ }^{1} \Pi_{1}$ & ${ }^{3} \Pi_{1}$ \\
 $4.03$ & $4.05$ & $4.05$ & ${ }^{1} \Sigma_{0}$ & ${ }^{1} \Sigma_{0}$ & ${ }^{1} \Sigma_{0}$ \\
 $4.07$ & $4.08$ & $4.09$ & ${ }^{3} \Sigma_{1}$ & ${ }^{3} \Sigma_{1}$ & ${ }^{3} \Sigma_{1}$ \\
 $4.26$ & $4.26$ & $4.26$ & ${ }^{3} \Delta_{3}$  &${ }^{3} \Delta_{3}$  & ${ }^{3} \Delta_{3}$\\
$4.36$ & $4.37$ & $4.37$ & ${ }^{1} \Delta_{2}$ & ${ }^{1} \Delta_{2}$ & ${ }^{1} \Delta_{2}$ \\
$4.73$ & $4.75$ & $4.75$ & ${ }^{3} \Sigma_{0}$ & ${ }^{3} \Sigma_{0}$ & ${ }^{3} \Sigma_{0}$ \\
\hline\hline
\end{tabular}
\end{center}                                                                    
\end{table}

\begin{table}[ht]                                                                   
\begin{center}                                                                  
\caption{Calibration of the number of holes in the O's $2p$-dominant spinors on the transition energy ($T_v(R);R=3.1609$ bohr). The number of holes corresponds to "$p$" in FIG 1 of ref. \cite{Fleig2016PRA_TaN}. MR$^{+S/D/T}_{10}$-CISD(8) model with virtual truncation at 2.5 ($E_H$) is employed.}\label{tbl:core}

\begin{tabular}{ccc|ccc}
\hline\hline
\multicolumn{3}{c}{$T_v$ (eV)} &\multicolumn{3}{c}{$^{2S+1}\Lambda_{\Omega}$} \\
1-hole & 2-hole & 3-hole  & 1-hole & 2-hole & 3-hole \\
\hline
 $0.00$ & $0.00$ & $0.00$  & ${ }^{3} \Delta_{1}$ & ${ }^{3} \Delta_{1}$ & ${ }^{3} \Delta_{1}$ \\
 $0.16$ & $0.16$ & $0.16$  & ${ }^{3} \Delta_{2}$ & ${ }^{3} \Delta_{2}$ & ${ }^{3} \Delta_{2}$ \\
 $0.42$ & $0.36$ & $0.39$  & ${ }^{3} \Delta_{3}$ & ${ }^{1} \Sigma_{0}$ & ${ }^{3} \Delta_{3}$ \\
 $0.46$ & $0.39$ & $0.41$  & ${ }^{1} \Sigma_{0}$ & ${ }^{3} \Delta_{3}$ & ${ }^{1} \Sigma_{0}$ \\
 $0.98$ & $1.16$ & $1.11$  & ${ }^{3} \Sigma_{0}$ & ${ }^{3} \Sigma_{0}$ & ${ }^{3} \Sigma_{0}$ \\
 $0.98$ & $1.23$ & $1.16$  & ${ }^{3} \Sigma_{1}$ & ${ }^{3} \Sigma_{1}$ & ${ }^{3} \Sigma_{1}$ \\
 $1.35$ & $1.34$ & $1.34$  & ${ }^{1} \Delta_{2}$ & ${ }^{1} \Delta_{2}$ & ${ }^{1} \Delta_{2}$ \\
 $1.61$ & $1.86$ & $1.80$  & ${ }^{1} \Gamma_{4}$ & ${ }^{1} \Gamma_{4}$ & ${ }^{1} \Gamma_{4}$ \\
 $1.80$ & $2.15$ & $2.23$  & ${ }^{3} \Phi_{2}$ & ${ }^{3} \Phi_{2}$ & ${ }^{3} \Phi_{2}$ \\
 $1.99$ & $2.17$ & $2.32$  & ${ }^{3} \Pi_{1}$ & ${ }^{3} \Pi_{0}$ & ${ }^{3} \Pi_{0}$ \\
 $2.04$ & $2.20$ & $2.34$  & ${ }^{3} \Pi_{0}$ & ${ }^{3} \Pi_{0}$ & ${ }^{3} \Pi_{0}$ \\
 $2.07$ & $2.23$ & $2.35$  & ${ }^{3} \Pi_{0}$ & ${ }^{3} \Pi_{1}$ & ${ }^{3} \Pi_{1}$ \\
 $2.19$ & $2.41$ & $2.42$  & ${ }^{3} \Phi_{3}$ & ${ }^{3} \Pi_{1}$ & ${ }^{1} \Sigma_{0}$ \\
 $2.20$ & $2.43$ & $2.54$  & ${ }^{3} \Pi_{1}$ & ${ }^{1} \Sigma_{0}$ & ${ }^{3} \Pi_{1}$ \\
 $2.30$ & $2.53$ & $2.62$  & ${ }^{1} \Sigma_{0}$ & ${ }^{3} \Phi_{3}$ & ${ }^{3} \Phi_{3}$ \\
 $2.39$ & $2.57$ & $2.72$  & ${ }^{3} \Pi_{2}$ & ${ }^{3} \Pi_{2}$ & ${ }^{3} \Pi_{2}$ \\
 $2.47$ & $2.79$ & $2.88$  & ${ }^{3} \Pi_{0}$ & ${ }^{3} \Pi_{0}$ & ${ }^{3} \Pi_{0}$ \\
 $2.50$ & $2.81$ & $2.89$  & ${ }^{3} \Pi_{0}$ & ${ }^{3} \Pi_{0}$ & ${ }^{3} \Pi_{0}$ \\
 $2.56$ & $2.83$ & $2.91$  & ${ }^{3} \Pi_{2}$ & ${ }^{3} \Pi_{2}$ & ${ }^{3} \Pi_{2}$ \\
 $2.60$ & $2.91$ & $3.02$  & ${ }^{3} \Phi_{4}$ & ${ }^{1} \Pi_{1}$ & ${ }^{3} \Phi_{4}$ \\
 $2.66$ & $2.93$ & $3.02$  & ${ }^{1} \Pi_{1}$ & ${ }^{3} \Phi_{4}$ & ${ }^{1} \Pi_{1}$ \\
 $3.26$ & $3.53$ & $3.63$  & ${ }^{1} \Phi_{3}$ & ${ }^{1} \Phi_{3}$ & ${ }^{1} \Phi_{3}$ \\
 $3.54$ & $3.74$ & $3.83$  & ${ }^{1} \Pi_{1}$ & ${ }^{3} \Delta_{1}$ & ${ }^{3} \Delta_{1}$ \\
 $3.73$ & $3.90$ & $3.97$  & ${ }^{3} \Delta_{1}$ & ${ }^{3} \Delta_{2}$ & ${ }^{3} \Delta_{2}$ \\
 $3.73$ & $3.92$ & $4.04$  & ${ }^{3} \Delta_{2}$ & ${ }^{1} \Pi_{1}$ & ${ }^{1} \Pi_{1}$ \\
 $4.10$ & $4.03$ & $4.17$  & ${ }^{3} \Delta_{3}$ & ${ }^{1} \Sigma_{0}$ & ${ }^{1} \Sigma_{0}$ \\
 $4.10$ & $4.07$ & $4.21$  & ${ }^{1} \Sigma_{0}$ & ${ }^{3} \Sigma_{1}$ & ${ }^{3} \Sigma_{1}$ \\
 $4.11$ & $4.26$ & $4.34$  & ${ }^{3} \Sigma_{1}$ & ${ }^{3} \Delta_{3}$ & ${ }^{3} \Delta_{3}$ \\
 $4.22$ & $4.36$ & $4.45$  & ${ }^{1} \Delta_{2}$ & ${ }^{1} \Delta_{2}$ & ${ }^{1} \Delta_{2}$ \\
 $4.36$ & $4.73$ & $4.89$  & ${ }^{3} \Sigma_{0}$ & ${ }^{3} \Sigma_{0}$ & ${ }^{3} \Sigma_{0}$ \\
\hline\hline
\end{tabular}
\end{center}                                                                    
\end{table}

\subsection{Spectroscopic constants}\label{sec:spec}
\begin{figure}[ht]
\begin{center}
\includegraphics[width=.5\textwidth]{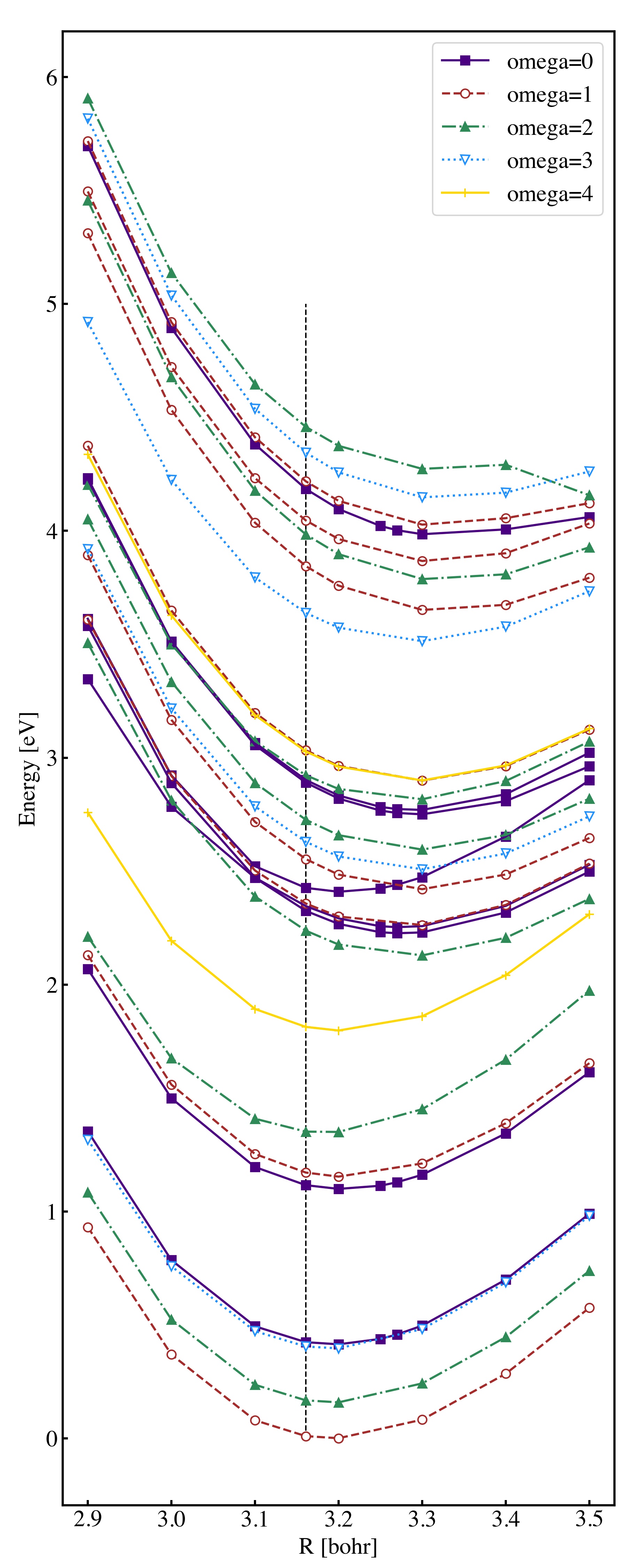}
  \caption{Potential energy curves for the twenty nine electronic state of TaO+, using the CI model $\mathrm{MR}_{10}^{+T}$-CISD(8), with cutoff 2.5 $E_\mathrm{H}$. The states shown in Table \ref{tbl:re_we_less} are described except for $^3 \Sigma_0$ ($T_v(R)$ = 4.889 eV). The energy offset is -15693.037948 $E_H$, which is the total electronic energy of the ground state ($^3 \Delta_1$) at the equilibrium structure (3.195 bohr). The vertical bar gives the internuclear distance (3.1609 bohr) used for vertical calculations shown in Tables \ref{tbl:active}-\ref{tbl:TDM_1.5} and S1 (supplemental material).
  }\label{fig:multiscale}
\end{center}
\end{figure}

The character of the electronic states can be seen from a higher perspective by summarizing  the spectroscopic constants and electronic configuration (Table \ref{tbl:re_we_less}) and the potential energy curves (PEC, FIG \ref{fig:multiscale}).
The PECs cannot be attributed to the atomic spectrum of Ta cation because of the strong mixing between Ta and O’s atomic spinors. The identification of $S$ and  $\Lambda$ in $^{2S+1} \Lambda_{\Omega}$ may be unreliable ($\Omega$, on the other hand, is a good quantum number) because of the contributions of a few pseudo-degenerate spinors. The magnitude of spin-orbit splittings is in many cases difficult to extract exactly due to configurational mixing in the involved states. However, we can make some remarks referring to the character of the spinors shown in Table \ref{tbl:MP}.

For the states with small $R_\mathrm{e}$ and large $\omega_\mathrm{e}$ (eight lowest electronic states and $^1 \Sigma_0$ ($T_v$ = 2.417 eV)), the electronic configuration mainly consists of $6s$ and $5d_\delta$ and almost purely of Ta’s atomic spinors. For the other states antibonding type ($5d_\pi$, $6p_\pi$, $6p_\sigma$) spinors make the bond length longer. 
For the high-energy region, there is an energetic gap between $^1 \Pi_1$ ($T_v$ = 3.024 eV) and $^1 \Phi_3$ ($T_v$ = 3.629 eV). However, the values of $R_\mathrm{e}$ and $\omega_\mathrm{e}$ do not vary significantly between the states above/below the gap. The reason for this is that the composition of the main electronic configuration of these states is similar: one electron occupies a lone-pair type spinor ($6s$ or $5d_\delta$) and another electron occupies an antibonding type spinor ($5d_\pi$, $6p_\pi$ or $6p_\sigma$). The whole spectrum is very dense, but no avoided crossings are to be found in this energy and internuclear separation range, except for the sixth excited state of $\Omega=2$ symmetry. 

In contrast to $R_\mathrm{e}$ and $\omega_\mathrm{e}$, the range of the static molecular dipole moments (PDM) is large. The minimum value (on the absolute) is $-0.494$ D for ${ }^{1} \Sigma_{0}$ ($T_v=4.175$ eV), and the maximum value is $-5.311$ D for ${ }^{3} \Sigma_{1}$ ($T_v=1.162$ eV). 
This trend can also be explained by the electronic configuration. The states with PDM greater than $-5.0$ D on the absolute mainly consist of $5d_\delta$, while in those where the PDM is smaller than $-1.1$ D the contribution of $6p_\sigma$ ($5d_\delta$) is large (small). The $5d_\delta$ molecular spinors are localized on the Ta nucleus because it is almost purely $5d_\delta$ atomic spinors. On the other hand, $6p_\sigma$ molecular spinors are of antibonding type, and their overlap with Oxygen’s spinors is larger than that of $\pi$-bonding-type molecular spinors. This leads to smaller electron density in the bonding region and larger electron density in the antibonding region ($z' \vec{e}_{z}$ region with $z'<0$ when the Ta nucleus is placed in the origin, and the O nucleus is placed at $z \vec{e}_{z}$ with $z
>0$, respectively). 

There are some excited states with large transition dipole moments (TDM) that were not addressed in the previous work \cite{PhysRevA.95.022504}. The excited states that have TDM greater than 1.5 eV are shown in Table \ref{tbl:TDM_1.5}, and a more detailed display is given in the supplemental material (Table S1). The sparse and off-diagonal character of the TDM matrix can be explained as follows: i) Although the electronic selection rules of diatomic molecules are in Hund's case c defined through $\Delta\Omega$, the TDM becomes largest when $\Delta S = 0$. In fact, when $\Delta S\neq0$ the transitions shown in Table \ref{tbl:TDM_1.5} produce a TDM less than $1.5$ D except for the transition between $^1\Sigma_0$ ($T_v=0.414$ eV) and $^3\Sigma_0$ ($T_v=4.889$ eV), and between $^1\Sigma_0$ ($T_v=0.414$ eV) and $^3\Pi_1$ ($T_v=2.348$ eV). ii) Large TDMs are found in the excitation from the states including ($6s$ or $5d_\delta$) spinors  to the states including ($5d_\pi$, $6p_\pi$, or $6p_\sigma$). The former states are low-energy states and correspond to the rows of the table, and the latter are high-energy states and correspond to the columns. When we focus on the ground state ($^3\Delta_1$), $6s\rightarrow6p_\sigma$ transition determines the large TDM of $^3\Delta_1$ ($T_v=3.835$ eV), while the excitation $6s\rightarrow5d_{\pi,1/2}$ and $5d_{\delta,3/2}\rightarrow5d_{\delta,5/2}$ occurs in ${ }^{3} \Phi_{2}$ state ($T_v$ = 2.228 eV). The Ta atomic $6p_{\pi,1/2}$ spinor also contributes in the $5d_{\pi,1/2}$-dominant molecular spinor (see no. 43 spinor in Table \ref{tbl:MP}) which is the reason for the large TDM in this state. Concerning $5d_{\delta,3/2}\rightarrow5d_{\delta,5/2}$, we should note that $\Delta l=\pm1$ is allowed in the atomic selection rule, while $\Delta \omega=0,\pm1$ is allowed in the molecular selection rule. 

\section{Conclusions}\label{sec:conclusion}
The present work is the first systematic and large-scale investigation of the electronic
structure of the TaO$^+$ cation which is a promising molecular species in the search for ${\cal{P}}$-
and ${\cal{P,T}}$-violating interactions \cite{PhysRevA.95.022504,Penyazkov2022CPL_PNC}.
We used an economic but sufficiently accurate relativistic CI model the reliability of which was confirmed through comparison with previous work \cite{PhysRevA.95.022504}. 

The electronic spectrum in an energy window of about
$5$ eV turns out to be dense, as anticipated. The potential-energy curves fall into two main categories which roughly correlate with the values of $R_\mathrm{e}$ and $\omega_\mathrm{e}$: i) States with two electrons in a lone-pair-like spinor ($6s$ or $5d_\delta$), and ii) States with one electron occupying the lone-pair-like spinor, and another occupying a ($5 d_{\pi}$, $6 p_{\pi}$, or $6 p_{\sigma}$) spinor. 
We, furthermore, present an extensive map of E1 transition moments within the $5$ eV energy window that should provide helpful information on how to access the target electronic science state. The largest E1 TDMs are found between states characterized by excitations from ($6s$ or $5d_\delta$) to ($5 d_{\pi}$, $6 p_{\pi}$, or $6 p_{\sigma}$) spinors.  

From a more general theoretical perspective, our present work is an extensive study on a large and encompassing number of excited states with relativistic CI theory. Similar calculations have been reported for diatomic or triatomic molecules, but in those cases the properties were limited to the vertical excitation energy \cite{Gomes2010JCP_I3-,Yamamoto2015JCP_DyF,Bross2015JCP_UF+}, or the employed active space was small \cite{Pototschnig2021PCCP_YbF}. 
On the experimental side, Resonant Enhanced Multiphoton Ionization (REMPI) is currently being used to create TaO$^+$ from its neutral precursor TaO in a supersonic cooled molecular beam \cite{Chung_Zhou2021}. Our present work should facilitate the identification of the TaO$^+$ cation and provide essential information for finding a pathway to creating the molecule in a desired state.


\begin{center}
\begin{longtable}{clccccc}
\caption{Characterization of the electronic ground and excited states (less than 5 eV). The wave-function model used is $\mathrm{MR}_{10}^{+T}$-CISD(8) with cutoff 2.5 $E_H$. The percentage is calculated as $100 \times|C_j|^2$ for a normalized CI wave function, where $j$ denotes an index of the Slater determinant with each configuration shown in the Table. Spectroscopic constants: $R_\mathrm {e}$, the equilibrium bond length; $\omega_\mathrm{e}$, the harmonic vibrational frequency; $T_v$, the excitation energy at $R$ = 3.1609 bohr; and PDM, molecule-frame static electric dipole moments (Debye) obtained at $R$, where O nucleus is placed at $z \vec{e}_{z}$ with $z>0$ (The coordinate origin for the PDM is the center of mass of TaO$^+$). Superscript $o$ indicates spinor occupation. Subscript $\lambda$ indicates the character of the bond in the non-relativistic regime. The values in parentheses refer to those in Ref. \cite{PhysRevA.95.022504}} \label{tbl:re_we_less}
\\
\hline
$^{2S+1}\Lambda_{\Omega}$ & atom$(nl_{\lambda,\omega})^o$ & $|C_j|^2$ & $R_\mathrm{e}$ (bohr) & $\omega_\mathrm{e}$ (cm$^{-1}$) & $T_v (\mathrm{eV})$ & $\mathrm{PDM}$(D) \\
\hline
 \endfirsthead
 \multicolumn{7}{l}{\small\it continued}\\
\hline
$^{2S+1}\Lambda_{\Omega}$ & atom$(nl_{\lambda,\omega})^o$ & $|C_j|^2$ & $R_\mathrm{e}$ (bohr) & $\omega_\mathrm{e}$ (cm$^{-1}$) & $T_v (\mathrm{eV})$ & $\mathrm{PDM}$(D) \\
\hline
 \endhead
 \multicolumn{7}{r}{\small\it continued on next page}\\
 \endfoot
\hline
 \multicolumn{7}{l}{ }\\
 \endlastfoot

${ }^{3} \Delta_{1}$ & $\mathrm{Ta}(6 s_{\sigma, 1 / 2})^{1}(5 d_{\delta, 3 / 2})^{1}$ & $0.90$ & $3.195$ & 1036 & 0.000 &  $ -4.036$\\
 &  &  & (3.161) & (1091) & (0.000) & $(-4.077)$\tabularnewline
 \hline${ }^{3} \Delta_{2}$ & $\mathrm{Ta}(6 s_{\sigma, 1 / 2})^{1}(5 d_{\delta, 3 / 2})^{1}$ & $0.61$ & $3.194$ & 1038 & $0.158$ & $ -4.002$ \\
& $\mathrm{Ta}(6 s_{\sigma, 1 / 2})^{1}(5 d_{\delta, 5 / 2})^{1}$ & $0.29$ & (3.160) & (1092) & (0.163) & $(-4.044)$\tabularnewline
\hline${ }^{3} \Delta_{3}$ & $\mathrm{Ta}(6 s_{\sigma, 1 / 2})^{1}(5 d_{\delta, 5 / 2})^{1}$ & $0.88$ & $3.193$ & 1039 & $0.395$ & $-4.006 $ \\
 &  &  & (3.160) & (1093) & (0.405) & $(-4.043)$\tabularnewline
\hline${ }^{1} \Sigma_{0}$ & $\mathrm{Ta}(6 s_{\sigma, 1 / 2})^{2}$ & $0.62$ & $3.194$ & 1039 & $0.414$ & $-3.826 $ \\
& $\mathrm{Ta}(5 d_{\delta, 3 / 2})^{2}$ & $0.22$ & (3.165) & (1086) & (0.466) & $(-4.109)$\tabularnewline
\hline${ }^{3} \Sigma_{0}$ & $\mathrm{Ta}(5 d_{\delta, 3 / 2})^{2}$ & $0.49$ & $3.206$ & 1013 & $1.106$ & $-5.064 $ \\
& $\mathrm{Ta}(5 d_{\delta, 5 / 2})^{2}$ & $0.30$ & (3.170) & (1071) & (1.025) & $(-4.960)$\tabularnewline
\hline${ }^{3} \Sigma_{1}$ & $\mathrm{Ta}(5 d_{\delta, 3 / 2})^{1}(5 d_{\delta, 5 / 2})^{1}$ & $0.88$ & $3.209$ & 1007 & $1.162$ & $-5.311 $\\
 &  &  & (3.174) & (1061) & (1.043) & $(-5.341)$\tabularnewline
\hline${ }^{1} \Delta_{2}$ & $\mathrm{Ta}(6 s_{\sigma, 1 / 2})^{1}(5 d_{\delta, 5 / 2})^{1}$ & $0.58$ & $3.182$ & 1047 & $1.342$ &$ -3.259$ \\
& $\mathrm{Ta}(6 s_{\sigma, 1 / 2})^{1}(5 d_{\delta, 3 / 2})^{1}$ & $0.26$ & (3.149) & (1101) & (1.421) & $(-3.309)$\tabularnewline

\hline${ }^{1} \Gamma_{4}$ & $\mathrm{Ta}(5 d_{\delta, 3 / 2})^{1}(5 d_{\delta, 5 / 2})^{1}$ & $0.88$ & $3.206$ & 1010 & $1.804$ & $-5.294 $ \\
\hline${ }^{3} \Phi_{2}$ & $\mathrm{Ta}(5 d_{\delta, 5 / 2})^{1}(5 d_{\pi, 1 / 2})^{1}$ & $0.88$ & $3.284$ & 929 & $2.228$ & $ -3.972$ \\
\hline${ }^{3} \Pi_{0}$ & $\mathrm{Ta}(6 s_{\sigma, 1 / 2})^{1}(6 p_{\pi, 1 / 2})^{1}$ & $0.44$ & $3.276$ & 946 & $2.315$ & $ -2.825$ \\
& $\mathrm{Ta}(6 s_{\sigma, 1 / 2})^{1}(6 p_{\pi, 1 / 2})^{1}$ & $0.44$ & & & & \\
\hline${ }^{3} \Pi_{0}$ & $\mathrm{Ta}(6 s_{\sigma, 1 / 2})^{1}(6 p_{\pi, 1 / 2})^{1}$ & $0.40$ & $3.274$ & 922 & $2.336$ &  $-2.951 $\\
& $\mathrm{Ta}(6 s_{\sigma, 1 / 2})^{1}(6 p_{\pi, 1 / 2})^{1}$ & $0.40$ & & & & \\

\hline${ }^{3} \Pi_{1}$ & $\mathrm{Ta}(6 s_{\sigma, 1 / 2})^{1}(5 d_{\pi, 1 / 2})^{1}$ & $0.62$ & $3.276$ & 942 & $2.348$ & $ -3.205$ \\
\hline${ }^{1} \Sigma_{0}$ & $\mathrm{Ta}(5 d_{\delta, 5 / 2})^{2}$ & $0.46$ & $3.205$ & 1030 & $2.417$ & $-4.209 $ \\
& $\mathrm{Ta}(6 s_{\sigma, 1 / 2})^{2}$ & $0.13$ & & & & \\
& $\mathrm{Ta}(5 d_{\delta, 3 / 2})^{2}$ & $0.13$ & & & & \\
\hline${ }^{3} \Pi_{1}$ & $\mathrm{Ta}(5 d_{\delta, 5 / 2})^{1}(5 d_{\pi, 3 / 2})^{1}$ & $0.66$ & $3.295$ & 927 & $2.543$ & $-3.480 $ \\
  & $\mathrm{Ta}(6 s_{\sigma, 1 / 2})^{1}(5 d_{\pi, 1 / 2})^{1}$ & $0.19$ & &  & &  \\
\hline${ }^{3} \Phi_{3}$ & $\mathrm{Ta}(5 d_{\delta, 5 / 2})^{1}(5 d_{\pi, 1 / 2})^{1}$ & $0.44$ & $3.290$ & 922 & $2.619$ & $-4.008 $ \\
& $\mathrm{Ta}(5 d_{\delta, 3 / 2})^{1}(5 d_{\pi, 3 / 2})^{1}$ & $0.44$ & & & & \\
\hline${ }^{3} \Pi_{2}$ & $\mathrm{Ta}(6 s_{\sigma, 1 / 2})^{1}(5 d_{\pi, 3 / 2})^{1}$ & $0.83$ & $3.295$ & 924 & $2.717$ &$-3.176 $ \\
\hline${ }^{3} \Pi_{0}$ & $\mathrm{Ta}(5 d_{\delta, 3 / 2})^{1}(6 p_{\pi, 3 / 2})^{1}$ & $0.44$ & $3.300$ & 916 & $2.880$ & $ -4.007$\\
& $\mathrm{Ta}(5 d_{\delta, 3 / 2})^{1}(6 p_{\pi, 3 / 2})^{1}$ & $0.44$ & & & & \\
\hline${ }^{3} \Pi_{0}$ & $\mathrm{Ta}(5 d_{\delta, 3 / 2})^{1}(6 p_{\pi, 3 / 2})^{1}$ & $0.44$ & $3.294$ & 939 & $2.892$ & $-3.989 $ \\
& $\mathrm{Ta}(5 d_{\delta, 3 / 2})^{1}(6 p_{\pi, 3 / 2})^{1}$ & $0.44$ & & & & \\
\hline${ }^{3} \Pi_{2}$ & $\mathrm{Ta}(5 d_{\delta, 5 / 2})^{1}(6 p_{\pi, 1 / 2})^{1}$ & $0.83$ & $3.282$ & 934 & $2.913$ & $-3.665 $\\
\hline${ }^{3} \Phi_{4}$ & $\mathrm{Ta}(5 d_{\delta, 5 / 2})^{1}(5 d_{\pi, 3 / 2})^{1}$ & $0.90$ & $3.294$ & 923 & $3.018$ & $-4.088 $ \\
\hline${ }^{1} \Pi_{1}$ & $\mathrm{Ta}(5 d_{\delta, 5 / 2})^{1}(5 d_{\pi, 3 / 2})^{1}$ & $0.45$ & $3.296$ & 927 & $3.024$ & $ -3.668$\\
& $\mathrm{Ta}(6 s_{\sigma, 1 / 2})^{1}(5 d_{\pi, 3 / 2})^{1}$ & $0.27$ & & & & \\
& $\mathrm{Ta}(5 d_{\delta, 3 / 2})^{1}(5 d_{\pi, 1 / 2})^{1}$ & $0.12$ & & & & \\
\hline${ }^{1} \Phi_{3}$ & $\mathrm{Ta}(5 d_{\delta, 5 / 2})^{1}(5 d_{\pi, 1 / 2})^{1}$ & $0.42$ & $3.294$ & 909 & $3.629$ & $ -3.542$ \\
& $\mathrm{Ta}(5 d_{\delta, 3 / 2})^{1}(5 d_{\pi, 3 / 2})^{1}$ & $0.42$ & & & & \\
\hline${ }^{3} \Delta_{1}$ & $\mathrm{Ta}(5 d_{\delta, 3 / 2})^{1}(6 p_{\sigma, 1 / 2})^{1}$ & $0.59$ & $3.329$ & 886 & $3.835$ & $ -2.081$ \\
& $\mathrm{Ta}(5 d_{\delta, 5 / 2})^{1}(5 d_{\pi, 3 / 2})^{1}$ & $0.15$ & & & & \\
\hline${ }^{3} \Delta_{2}$ & $\mathrm{Ta}(5 d_{\delta, 3 / 2})^{1}(6 p_{\sigma, 1 / 2})^{1}$ & $0.81$ & $3.331$ & 889 & $3.974$ & $ -1.601$\\
\hline${ }^{1} \Pi_{1}$ & $\mathrm{Ta}(5 d_{\delta, 3 / 2})^{1}(6 p_{\sigma, 1 / 2})^{1}$ & $0.27$ & $3.320$ & 902 & $4.036$ &$ -2.598$ \\
& $\mathrm{Ta}(5 d_{\pi, 3 / 2})^{1}(5 d_{\delta, 5 / 2})^{1}$ & $0.19$ & & & & \\
& $\mathrm{Ta}(6 s_{\sigma, 1 / 2})^{1}(5 d_{\pi, 3 / 2})^{1}$ & $0.12$ & & & \\
& $\mathrm{Ta}(6 s_{\sigma, 1 / 2})^{1}(6 p_{\sigma, 1 / 2})^{1}$ & $0.12$ & & & \\

\hline${ }^{1} \Sigma_{0}$ & $\mathrm{Ta}(6 s_{\sigma, 1 / 2})^{1}(6 p_{\sigma, 1 / 2})^{1}$ & $0.44$ & $3.330$ & 899 & $4.175$ & $-0.494 $\\
& $\mathrm{Ta}(6 s_{\sigma, 1 / 2})^{1}(6 p_{\sigma, 1 / 2})^{1}$ & $0.44$ & & & & \\
\hline${ }^{3} \Sigma_{1}$ & $\mathrm{Ta}(6 s_{\sigma, 1 / 2})^{1}(6 p_{\sigma, 1 / 2})^{1}$ & $0.74$ & $-$ & $-$ & $4.209$ &$-1.098 $ \\
\hline${ }^{3} \Delta_{3}$ & $\mathrm{Ta}(5 d_{\delta, 5 / 2})^{1}(6 p_{\sigma, 1 / 2})^{1}$ & $0.86$ & $-$ & $-$ & $4.335$ & $ -1.486$ \\
\hline${ }^{1} \Delta_{2}$ & $\mathrm{Ta}(5 d_{\delta, 5 / 2})^{1}(6 p_{\sigma, 1 / 2})^{1}$ & $0.76$ & $-$ & $-$ & $4.449$ & $-1.877 $\\
\hline${ }^{3} \Sigma_{0}$ & $\mathrm{Ta}(6 s_{\sigma, 1 / 2})^{1}(6 p_{\sigma, 1 / 2})^{1}$ & $0.28$ & $-$ & $-$ & $4.889$ & $ -2.179$ \\
& $\mathrm{Ta}(6 s_{\sigma, 1 / 2})^{1}(6 p_{\sigma, 1 / 2})^{1}$ & $0.28$ & & & & \\
& $\mathrm{Ta}(5 d_{\pi, 1 / 2})^{2}$ & $0.18$ & & & & \\

\end{longtable}
\end{center}
                                  
%

\begin{table}
\begin{center}
\caption{Transition dipole moments (Debye) $\left\|\left\langle{ }^{M^{\prime}} \Lambda^{\prime}_{\Omega^{\prime}}|\vec{D_z}|^{M} \Lambda_{\Omega}\right\rangle\right\|$, with $\vec{D_z}$ the electric dipole moment operator, using the CI model $\mathrm{MR}_{10}^{+T}$-CISD(8), with cutoff 2.5 $E_H$. The bond length is $R=3.1609 $ bohr. Matrix elements smaller than 1.5 Debye are not shown.}\label{tbl:TDM_1.5}
\begin{tabular}{cc|cccccc}
\hline & $T_v(R)$ (eV) & $0.000$ & $0.158$ & $0.395$ & $0.414$ & $1.342$ & $1.804$ \\
$T_v(R)$ (eV) & term & ${ }^{3} \Delta_{1}$ & ${ }^{3} \Delta_{2}$ & ${ }^{3} \Delta_{3}$ & ${ }^{1} \Sigma_{0}$ & ${ }^{1} \Delta_{2}$ & ${ }^{1} \Gamma_{4}$ \\
\hline $2.228$ & ${ }^{3} \Phi_{2}$ & $2.448$ & & & & & \\
$2.348$ & ${ }^{3} \Pi_{1}$ & & & & $1.722$ & & \\
$2.543$ & ${ }^{3} \Pi_{1}$ & & $2.027$ & & & & \\
$2.619$ & ${ }^{3} \Phi_{3}$ & & $2.301$ & & & & \\
$2.717$ & ${ }^{3} \Pi_{2}$ & & & $1.526$ & & & \\
$2.913$ & ${ }^{3} \Pi_{2}$ & & & $1.855$ & & & \\
$3.018$ & ${ }^{3} \Phi_{4}$ & & & $2.220$ & & & \\
$3.629$ & ${ }^{1} \Phi_{3}$ & & & & & $1.932$ & $1.962$ \\
$3.835$ & ${ }^{3} \Delta_{1}$ & $2.035$ & & & & & \\
$3.974$ & ${ }^{3} \Delta_{2}$ & & $2.275$ & & & & \\
$4.335$ & ${ }^{3} \Delta_{3}$ & & & $2.316$ & & & \\
$4.449$ & ${ }^{1} \Delta_{2}$ & & & & & $2.410$ & \\
$4.889$ & ${ }^{3} \Sigma_{0}$ & & & & $1.872$ & & \\
\hline
\end{tabular}
\end{center}
\end{table}

\section{Acknowledgements}
We thank Dr. Andrew Jayich (Santa Barbara) and Dr. Yan Zhou (Las Vegas) for helpful discussions. A.S. acknowledges financial support from the Japan Society for the Promotion of Science (JSPS) KAKENHI (Grant No. 17J02767, 20K22553, and 21K14643), and JSPS Overseas Challenge Program for Young Researchers, Grant No. 201880193.


\bibliography{article,all}

\newcommand{\Aa}[0]{Aa}
\begin{thebibliography}{10}
\expandafter\ifx\csname url\endcsname\relax
  \def\url#1{\texttt{#1}}\fi
\expandafter\ifx\csname urlprefix\endcsname\relax\def\urlprefix{URL }\fi
\expandafter\ifx\csname href\endcsname\relax
  \def\href#1#2{#2} \def\path#1{#1}\fi

\bibitem{Kobayashi}
M.~Kobayashi, T.~Maskawa, {CP violation in the renormalizable theory of weak
  interaction}, Prog. Theor. Phys. 49 (1973) 652.

\bibitem{HfF+_EDM_PRL2017}
W.~B. Cairncross, D.~N. Gresh, M.~Grau, K.~C. Cossel, T.~S. Roussy, Y.~Ni,
  Y.~Zhou, J.~Ye, E.~A. Cornell, Precision measurement of the electron’s
  electric dipole moment using trapped molecular ions, Phys. Rev. Lett. 119
  (2017) 153001.

\bibitem{Cornell_TrapEDM_2020}
Y.~Zhou, Y.~Shagam, W.~B. Cairncross, K.~B. Ng, T.~S. Roussy, T.~Grogan,
  K.~Boyce, A.~Vigil, M.~Pettine, T.~Zelevinsky, J.~Ye, E.~A. Cornell,
  {Second-Scale Coherence Measured at the Quantum Projection Noise Limit with
  Hundreds of Molecular Ions}, Phys. Rev. Lett. 124 (2020) 053201.

\bibitem{ACME_ThO_eEDM_nature2018}
{Improved limit on the electric dipole moment of the electron}, Nature 562
  (2018) 355, {ACME} {C}ollaboration.

\bibitem{hudson_hinds_YbF2011}
J.~J. Hudson, D.~M. Kara, I.~J. Smallman, B.~E. Sauer, M.~R. Tarbutt, E.~A.
  Hinds, Improved measurement of the shape of the electron, Nature 473 (2011)
  493.

\bibitem{Sushkov_Flambaum_Khriplovich1984}
O.~P. Sushkov, V.~V. Flambaum, I.~B. Khriplovich, Possibility of investigating
  ${P}$- and ${T}$-odd nuclear forces in atomic and molecular experiments, Sov.
  Phys. JETP 60 (1984) 873.

\bibitem{Veneziano_Witten_QCD-PTodd1980}
R.~J. Crewther, P.~di~Vecchia, G.~Veneziano, E.~Witten, Chiral estimate of the
  electric dipole moment of the neutron in quantum chromodynamics, Phys. Lett.
  B 88 (1979) 123.

\bibitem{Gunion_Wyler_CEDM_NEDM_1990}
J.~F. Gunion, D.~Wyler, Inducing a large neutron electric dipole moment via a
  quark chromo-electric dipole moment, Phys. Lett. B 248 (1990) 170.

\bibitem{PhysRevA.95.022504}
T.~Fleig,
  \href{http://link.aps.org/doi/10.1103/PhysRevA.95.022504}{{T}a{O}$^{+}$ as a
  candidate molecular ion for searches of physics beyond the standard model},
  Phys. Rev. A 95 (2017) 022504.
\newblock \href {https://doi.org/10.1103/PhysRevA.95.022504}
  {\path{doi:10.1103/PhysRevA.95.022504}}.
\newline\urlprefix\url{http://link.aps.org/doi/10.1103/PhysRevA.95.022504}

\bibitem{Zhou2019JMS_ThF+}
Y.~Zhou, K.~B. Ng, L.~Cheng, D.~N. Gresh, R.~W. Field, J.~Ye, E.~A. Cornell,
  {Visible and ultraviolet laser spectroscopy of ThF}, J. Mol. Spectrosc. 358
  (2019) 1--16.
\newblock \href {http://arxiv.org/abs/1901.06084} {\path{arXiv:1901.06084}},
  \href {https://doi.org/10.1016/j.jms.2019.02.006}
  {\path{doi:10.1016/j.jms.2019.02.006}}.

\bibitem{cornell_ThF+_BAPS_2021}
N.~Schlossberger, K.~B. Ng, S.~Y. Park, Y.~Zhou, T.~Roussy, T.~Grogan,
  Y.~Shagam, A.~Vigil, M.~Pettine, J.~Ye, E.~A. Cornell, {Spectroscopy of
  ThF$^+$ in aim of a new eEDM measurement with trapped molecular ions},
  Bulletin of the American Physical Society 65 (2020/6/3).

\bibitem{Ng2022PRA_ThF+}
K.~B. Ng, Y.~Zhou, L.~Cheng, N.~Schlossberger, S.~Y. Park, T.~S. Roussy,
  L.~Caldwell, Y.~Shagam, A.~J. Vigil, E.~A. Cornell, J.~Ye, Spectroscopy on
  the electron-electric-dipole-moment--sensitive states of {ThF$^{+}$}, Phys.
  Rev. A 105~(2) (2022) 022823.
\newblock \href {https://doi.org/10.1103/PhysRevA.105.022823}
  {\path{doi:10.1103/PhysRevA.105.022823}}.

\bibitem{ThF+_NJP_2015}
M.~Denis, M.~N{\o}rby, H.~J.~{\Aa}. Jensen, A.~S.~P. Gomes, M.~K. Nayak,
  S.~Knecht, T.~Fleig, Theoretical study on {T}h{F}$^+$, a prospective system
  in search of time-reversal violation, New J. Phys. 17 (2015) 043005.

\bibitem{Gresh2016JMS}
D.~N. Gresh, K.~C. Cossel, Y.~Zhou, J.~Ye, E.~A. Cornell, {Broadband velocity
  modulation spectroscopy of ThF+ for use in a measurement of the electron
  electric dipole moment}, J. Mol. Spectrosc. 319 (2016) 1--9.
\newblock \href {http://arxiv.org/abs/1509.03682} {\path{arXiv:1509.03682}},
  \href {https://doi.org/10.1016/j.jms.2015.11.001}
  {\path{doi:10.1016/j.jms.2015.11.001}}.

\bibitem{Cairncross_Ye_NatPhys2019}
W.~B. Cairncross, J.~Ye, {Atoms and molecules in the search for time-reversal
  symmetry violation}, Nature Physics 1 (2019) 510.

\bibitem{Flambaum_DeMille_Kozlov2014}
V.~V. Flambaum, D.~DeMille, M.~G. Kozlov, Time-{R}eversal {S}ymmetry
  {V}iolation in {M}olecules {I}nduced by {N}uclear {M}agnetic {Q}uadrupole
  {M}oments, Phys. Rev. Lett. 113 (2014) 103003.

\bibitem{Lackenby2018PRA_MQM}
B.~G. Lackenby, V.~V. Flambaum, {Time reversal violating magnetic quadrupole
  moment in heavy deformed nuclei}, Phys. Rev. D 98~(11) (2018) 115019.
\newblock \href {http://arxiv.org/abs/1810.02477} {\path{arXiv:1810.02477}},
  \href {https://doi.org/10.1103/PhysRevD.98.115019}
  {\path{doi:10.1103/PhysRevD.98.115019}}.

\bibitem{Penyazkov2022CPL_PNC}
G.~Penyazkov, L.~V. Skripnikov, A.~V. Oleynichenko, A.~V. Zaitsevskii, Effect
  of the neutron quadrupole distribution in the {TaO+} cation, Chem. Phys.
  Lett. 793 (2022) 139448.
\newblock \href {https://doi.org/10.1016/j.cplett.2022.139448}
  {\path{doi:10.1016/j.cplett.2022.139448}}.

\bibitem{CarrDeMille_NJP2009}
L.~D. Carr, D.~DeMille, R.~V. Krems, J.~Ye, Cold and ultracold molecules:
  science, technology and applications, New J. Phys. 11 (2009) 055049.

\bibitem{Cornell_MolIons_Science2013}
H.~Loh, K.~Cossel, M.~C. Grau, K.-K. Ni, E.~R. Meyer, J.~L. Bohn, J.~Ye, E.~A.
  Cornell, Precision {S}pectroscopy of {P}olarized {M}olecules in an {I}on
  {T}rap, Science 342 (2013) 1220.

\bibitem{Ni_HfF+_JMS2014}
K.-K. Ni, H.~Loh, M.~Grau, K.~C. Cossel, J.~Ye, E.~A. Cornell, State-specific
  detection of trapped {H}f{F}$^+$ by photoassociation, J. Mol. Spectrosc. 300
  (2014) 12.

\bibitem{knecht_luciparII}
S.~Knecht, H.~J.~A. Jensen, T.~Fleig, {L}arge-{S}cale {P}arallel
  {C}onfiguration {I}nteraction. {II}. {T}wo- and four-component double-group
  general active space implementation with application to {B}i{H}, J. Chem.
  Phys. 132 (2010) 014108.
\newblock \href {https://doi.org/10.1063/1.3276157}
  {\path{doi:10.1063/1.3276157}}.

\bibitem{fleig_gasci2}
T.~Fleig, J.~Olsen, L.~Visscher, The generalized active space concept for the
  relativistic treatment of electron correlation. {II}: {L}arge-scale
  configuration interaction implementation based on relativistic 2- and
  4-spinors and its application, J. Chem. Phys. 119 (2003) 2963.

\bibitem{DIRAC15}
DIRAC, a relativistic ab initio electronic structure program, Release DIRAC15
  (2015), written by R. Bast, T. Saue, L. Visscher, and H. J. Aa. Jensen, with
  contributions from V. Bakken, K. G. Dyall, S. Dubillard, U. Ekstroem, E.
  Eliav, T. Enevoldsen, E. Fasshauer, T. Fleig, O. Fossgaard, A. S. P. Gomes,
  T. Helgaker, J. Henriksson, M. Ilias, Ch. R. Jacob, S. Knecht, S. Komorovsky,
  O. Kullie, J. K. Laerdahl, C. V. Larsen, Y. S. Lee, H. S. Nataraj, M. K.
  Nayak, P. Norman, G. Olejniczak, J. Olsen, Y. C. Park, J. K. Pedersen, M.
  Pernpointner, R. Di Remigio, K. Ruud, P. Salek, B. Schimmelpfennig, J.
  Sikkema, A. J. Thorvaldsen, J. Thyssen, J. van Stralen, S. Villaume, O.
  Visser, T. Winther, and S. Yamamoto (see http://www.diracprogram.org).

\bibitem{Saue2020JCP}
T.~Saue, R.~Bast, A.~S.~P. Gomes, H.~J.~A. Jensen, L.~Visscher, I.~A. Aucar,
  R.~{Di Remigio}, K.~G. Dyall, E.~Eliav, E.~Fasshauer, T.~Fleig, L.~Halbert,
  E.~D. Hedeg{\aa}rd, B.~Helmich-Paris, M.~Ilia{\v{s}}, C.~R. Jacob, S.~Knecht,
  J.~K. Laerdahl, M.~L. Vidal, M.~K. Nayak, M.~Olejniczak, J.~M.~H. Olsen,
  M.~Pernpointner, B.~Senjean, A.~Shee, A.~Sunaga, J.~N.~P. van Stralen, {The
  DIRAC code for relativistic molecular calculations}, J. Chem. Phys. 152~(20)
  (2020) 204104.
\newblock \href {http://arxiv.org/abs/2002.06121} {\path{arXiv:2002.06121}},
  \href {https://doi.org/10.1063/5.0004844} {\path{doi:10.1063/5.0004844}}.

\bibitem{Visscher1997TCA}
L.~Visscher, {Approximate molecular relativistic Dirac-Coulomb calculations
  using a simple Coulombic correction}, Theor. Chem. Accounts Theory, Comput.
  Model. (Theoretica Chim. Acta) 98~(2-3) (1997) 68--70.
\newblock \href {https://doi.org/10.1007/s002140050280}
  {\path{doi:10.1007/s002140050280}}.

\bibitem{Dyall2004TCA_5d}
K.~G. Dyall, {Relativistic double-zeta, triple-zeta, and quadruple-zeta basis
  sets for the 5d elements Hf?Hg}, Theor. Chem. Acc. 112~(5-6) (2004) 403--409.
\newblock \href {https://doi.org/10.1007/s00214-004-0607-y}
  {\path{doi:10.1007/s00214-004-0607-y}}.

\bibitem{Dyall2009TCA_5d}
K.~G. Dyall, A.~S. Gomes, {Revised relativistic basis sets for the 5d elements
  Hf-Hg}, Theor. Chem. Acc. 125~(1) (2010) 97--100.
\newblock \href {https://doi.org/10.1007/s00214-009-0717-7}
  {\path{doi:10.1007/s00214-009-0717-7}}.

\bibitem{Dyall2016TCA_1-3}
K.~G. Dyall, {Relativistic double-zeta, triple-zeta, and quadruple-zeta basis
  sets for the light elements H–Ar}, Theor. Chem. Acc. 135~(5) (2016) 128.
\newblock \href {https://doi.org/10.1007/s00214-016-1884-y}
  {\path{doi:10.1007/s00214-016-1884-y}}.

\bibitem{Stanton1984JCP}
R.~E. Stanton, S.~Havriliak, {Kinetic balance: A partial solution to the
  problem of variational safety in Dirac calculations}, J. Chem. Phys. 81~(4)
  (1984) 1910--1918.
\newblock \href {https://doi.org/10.1063/1.447865}
  {\path{doi:10.1063/1.447865}}.

\bibitem{Visscher1997ADNDT}
L.~Visscher, K.~G. Dyall, {Dirac – Fock Atomic Electronic Structure
  Calculations Using Different Nuclear Charge Distributions}, At. Data Nucl.
  Data Tables 67~(2) (1997) 207--224.

\bibitem{Fleig2003JCP}
T.~Fleig, J.~Olsen, L.~Visscher, {The generalized active space concept for the
  relativistic treatment of electron correlation. II. Large-scale configuration
  interaction implementation based on relativistic 2- and 4-spinors and its
  application}, J. Chem. Phys. 119~(6) (2003) 2963--2971.
\newblock \href {https://doi.org/10.1063/1.1590636}
  {\path{doi:10.1063/1.1590636}}.

\bibitem{Knecht2010JCP}
S.~Knecht, H.~J.~A. Jensen, T.~Fleig, {Large-scale parallel configuration
  interaction. II. Two- and four-component double-group general active space
  implementation with application to BiH}, J. Chem. Phys. 132~(1) (2010)
  014108.
\newblock \href {https://doi.org/10.1063/1.3276157}
  {\path{doi:10.1063/1.3276157}}.

\bibitem{Knecht_thesis}
Stefan R. Knecht. Parallel Relativistic Multiconfiguration Methods: New
  Powerful Tools for Heavy-Element Electronic-Structure Studies. PhD thesis,
  Mathematisch-Naturwissenschaftliche Fakult{\"{a}}t,
  Heinrich-Heine-Universit{\"{a}}t D{\"{u}}sseldorf, 2009. URL:
  http://docserv.uni-duesseldorf.de/servlets/DocumentServlet?id=13226.

\bibitem{Denis2015NJP}
M.~Denis, M.~S. Norby, H.~J.~A. Jensen, A.~S.~P. Gomes, M.~K. Nayak, S.~Knecht,
  T.~Fleig, {Theoretical study on ThF+, a prospective system in search of
  time-reversal violation}, New J. Phys. 17 (2015).
\newblock \href {https://doi.org/10.1088/1367-2630/17/4/043005}
  {\path{doi:10.1088/1367-2630/17/4/043005}}.

\bibitem{Fleig2016PRA_TaN}
T.~Fleig, M.~K. Nayak, M.~G. Kozlov, {TaN, a molecular system for probing
  <math> <mrow> <mi mathvariant="script">P</mi> <mo>,</mo> <mi
  mathvariant="script">T</mi> </mrow> </math> -violating hadron physics}, Phys.
  Rev. A 93~(1) (2016) 012505.
\newblock \href {https://doi.org/10.1103/PhysRevA.93.012505}
  {\path{doi:10.1103/PhysRevA.93.012505}}.

\bibitem{Thyssen_thesis}
Jørn Thyssen. Development and Applications of Methods for Correlated
  Relativistic Calculations of Molecular Properties. PhD thesis, University of
  Southern Denmark, 2001. URL:
  http://dirac.chem.sdu.dk/thesis/thesis-jth2001.pdf.

\bibitem{fleig_nayak_eEDM2013}
T.~Fleig, M.~K. Nayak, Electron electric-dipole-moment interaction constant for
  {H}f{F}$^+$ from relativistic correlated all-electron theory, Phys. Rev. A 88
  (2013) 032514.

\bibitem{Denis-Fleig_ThO_JCP2016}
M.~Denis, T.~Fleig, In search of discrete symmetry violations beyond the
  standard model: {T}horium monoxide reloaded, J. Chem. Phys. 145 (2016)
  028645.

\bibitem{zenodo:dataset}
A.~Sunaga, T.~Fleig, {Spectroscopic and Electric Properties of the TaO$^+$
  Molecule Ion for the Search of New Physics: A Platform for Identification and
  State Control: Dataset (Version 1.0)}, Zenodo
  \url{http://doi.org/10.5281/zenodo.6339977} (2022).
\newblock \href {https://doi.org/10.5281/zenodo.6339977}
  {\path{doi:10.5281/zenodo.6339977}}.

\bibitem{Gomes2010JCP_I3-}
A.~S.~P. Gomes, L.~Visscher, H.~Bolvin, T.~Saue, S.~Knecht, T.~Fleig, E.~Eliav,
  The electronic structure of the triiodide ion from relativistic correlated
  calculations: a comparison of different methodologies, J. Chem. Phys. 133~(6)
  (2010) 064305.
\newblock \href {https://doi.org/10.1063/1.3474571}
  {\path{doi:10.1063/1.3474571}}.

\bibitem{Yamamoto2015JCP_DyF}
S.~Yamamoto, H.~Tatewaki, Electronic spectra of {DyF} studied by four-component
  relativistic configuration interaction methods, J. Chem. Phys. 142~(9) (2015)
  094312.
\newblock \href {https://doi.org/10.1063/1.4913631}
  {\path{doi:10.1063/1.4913631}}.

\bibitem{Bross2015JCP_UF+}
D.~H. Bross, K.~A. Peterson, Theoretical spectroscopy study of the low-lying
  electronic states of {UX} and {UX(+)}, {X} = {F} and cl, J. Chem. Phys.
  143~(18) (2015) 184313.
\newblock \href {https://doi.org/10.1063/1.4935492}
  {\path{doi:10.1063/1.4935492}}.

\bibitem{Pototschnig2021PCCP_YbF}
J.~V. Pototschnig, K.~G. Dyall, L.~Visscher, A.~S.~P. Gomes, Electronic spectra
  of ytterbium fluoride from relativistic electronic structure calculations,
  Phys. Chem. Chem. Phys. 23~(39) (2021) 22330--22343.
\newblock \href {https://doi.org/10.1039/d1cp03701c}
  {\path{doi:10.1039/d1cp03701c}}.

\bibitem{Chung_Zhou2021}
T.~Chung, M.~C. Cooper, Y.~Zhou, Tantalum oxide spectroscopy to facilitate
  exploring New Physics beyond the Standard Model, 2021 International Symposium
  on Molecular Spectroscopy,
  {\em{http://isms.illinois.edu/2021/schedule/2021\_Abstract\_Book.pdf}}.

\end{thebibliography}

\end{document}